\pdfoutput=1
\documentclass{article}

\usepackage{jcappub}

\usepackage[utf8]{inputenc}

\usepackage{hyperref}

\usepackage{amsmath}
\usepackage{amssymb}
\usepackage{xcolor}
\usepackage{physics}
\usepackage{graphicx} 

\usepackage{pgf} 

\numberwithin{equation}{section}

\begin{document}

\title{On the Wormhole--Warp Drive Correspondence}
\author{Remo Garattini}
\emailAdd{remo.garattini@unibg.it}
\author{and Kirill Zatrimaylov}
\emailAdd{kirill.zatrimaylov@sns.it, kirill.zatrimaylov@guest.unibg.it}
\affiliation{\emph{Università degli Studi di Bergamo, Dipartimento di Ingegneria e Scienze Applicate}\\\emph{Viale Marconi 5, 24044, Dalmine (Bergamo), Italy}}

\abstract{We propose a correspondence between the Morris--Thorne wormhole metric and a warp drive metric, which generalizes an earlier result by H. Ellis regarding the Schwarzschild black hole metric and makes it possible to embed a warp drive in a wormhole background. We demonstrate that in order to do that, one needs to also generalize the Natario--Alcubierre definition of warp drive and introduce nonzero intrinsic curvature. However, we also find out that in order to be traversable by a warp drive, the wormhole should have a horizon: in other words, humanly traversable wormholes cannot be traversed by a warp drive, and vice versa. We also discuss possible loopholes in this "no-go" theorem.}

\maketitle

\section{Introduction}\label{sec:introduction}
A warp drive, as defined by Miguel Alcubierre in~\cite{Alcubierre:1994tu}, is a particular solution of General Relativity that appears as a localized, bubble--shaped distortion of spacetime. The warp bubble can host a spacecraft and propagate through space at any speed (in principle including superluminal), but in order to create it, one needs negative energy density~\cite{Natario:2001tk}--\cite{Alcubierre:2017kqf}.

As shown by Ellis in~\cite{Ellis:2004aw}, the Schwarzschild metric can be represented as a warp spacetime with the use of Painlevé--Gullstrand coordinates, which makes it possible to consider a warp drive in a black hole background. In this paper, we generalize this result to the case of Morris--Thorne wormholes, a yet another solution of GR that requires negative energy density~\cite{Ellis:1973yv,Bronnikov:1973fh,Morris:1988cz}. Namely, in section~\ref{sec:Correspondence} we consider the wormhole traversability conditions and prove that the combination of the flare--out condition and the no--horizon condition inevitably produces a singularity. While this singularity is just a coordinate one for an ordinary observer, it would effectively be a real gravitational singularity for the warp drive. We also find that it may be possible to avoid it by considering a coordinate transformation that generalizes the notion of proper length; however, it comes at the cost of introducing new mathematical degrees of freedom without an immediate physical meaning. We conclude in section~\ref{sec:conclusions} with an overview of this result and some possible ways to circumvent it.

\section{The Wormhole--Warp Drive Correspondence}\label{sec:Correspondence}
A warp drive metric, as defined by Natario in~\cite{Natario:2001tk}, is given by
\begin{equation}
\label{Natario}
-dt^2+\sum^3_{i=1}(dx^i+N^i(\vec{r},t)dt)^2 \ .
\end{equation}
In ADM variables, this corresponds to the lapse function $N$ equal to $1$, and the inner metric $h_{ij}$ equal to $\delta_{ij}$, while $N^i$ is the shift vector.

The warp drive itself is a deformation of the metric that is localized in a bubble--shaped region moving on some (flat or non--flat) spacetime background with the velocity
\begin{equation}
\vec{v}_s(t)=\frac{dr_s}{dt} \ ,
\end{equation}
where $r_s(t)$ is the positon of the warp bubble's center.

This means that the functions $N^i$ have the form
\begin{equation}
N^i \ = \ (1-f(\vec{r},t))N^i_{out}(\vec{r},t)+f(\vec{r},t)N^i_{in}(t) \ ,
\end{equation}
where $N^i_{out}(\vec{r},t)$ is the background metric, $N^i_{in}(t)$ is the perturbation, and $f(|\vec{r}-\vec{r}_s(t)|)$ is a bell--shaped function describing the shape of the bubble.

In the particular case when the background metric is spherically symmetric, $N^i_{out}$ are given by
\begin{equation}
N^i_{out}(\vec{r},t) \ = \ \beta(\vec{r},t)\frac{x^i}{r} \ .
\end{equation}

In this case, the background metric can also be written in the more compact form in spherical coordinates
\begin{equation}\label{C1}
-dt^2+(dr+\beta(\vec{r},t)dt)^2+r^2d\Omega^2 \ .
\end{equation}
As found by Painlevé and Gullstrand, the Schwarzshild metric
\begin{equation}
-(1-\frac{2GM}{r})dt^2+\frac{dr^2}{1-\frac{2GM}{r}}+r^2d\Omega^2
\end{equation}
can be brought to the form~\eqref{C1} with 
\begin{equation}
\beta \ = \ \sqrt{\frac{2GM}{r}}
\end{equation}
via a coordinate transformation
\begin{equation}\label{PG}
t \ = \ T \ - \ \int \ dr \ \frac{\sqrt{\frac{2GM}{r}}}{1-\frac{2GM}{r}} \ .
\end{equation}
As suggested by Ellis in~\cite{Ellis:2004aw}, this relation can be used to embed an actual warp drive within the exterior of a black hole by replacing
\begin{equation}
N^i \ = \ \beta\frac{x^i}{r} \rightarrow (1-f(\vec{r},t))\beta\frac{x^i}{r}-f(\vec{r},t)v^i_{in}(t) \ .
\end{equation}
In this paper, we generalize the Painlevé--Gullstrand coordinates to Morris--Thorne wormholes, with the metric
\begin{equation}
-e^{2\Phi(r)}dt^2 \ + \ \frac{dr^2}{1-\frac{b(r)}{r}} \ + \ r^2d\Omega^2 \ .
\end{equation}
A generic coordinate transformation has the form
\begin{equation}
dt \ = \ \xi dT \ + \ \eta dr \ ;
\end{equation}
however, since our metric background is time independent, then, given the condition 
\begin{equation}
\partial_r\xi \ = \ \partial_T\eta \ ,
\end{equation}
$\xi$ has to be a constant that can be set to $1$ by rescaling.

After the transformation, the line element would be given by
\begin{equation}
-dT^2 \ + \ (1-e^{2\Phi(r)})dT^2 \ - \ 2e^{2\Phi(r)}\eta(r)dTdr \ + \ \left(\frac{1}{1-\frac{b(r)}{r}}-\eta^2(r)e^{2\Phi(r)}\right)dr^2 \ + \ r^2d\Omega^2 \ .
\end{equation}

In order for the second, the third, and the fourth term to comprise a full square, we have to choose
\begin{equation}
\eta(r) \ = \ -\frac{\sqrt{e^{-2\Phi(r)}-1}}{\sqrt{1-\frac{b(r)}{r}}} \ .
\end{equation}
The transformation we seek is therefore given by
\begin{equation}\label{T}
t \ = \ T \ - \ \int \ dr \ \frac{\sqrt{e^{-2\Phi(r)}-1}}{\sqrt{1-\frac{b(r)}{r}}} \ ,
\end{equation}
and it allows to represent the Morris--Thorne metric as
\begin{equation}\label{E}
-dT^2+(g(r)dr+\beta(r)dT)^2+r^2d\Omega^2 \ ,
\end{equation}
with 
\begin{equation}\label{B}
\beta \ = \ \sqrt{1-e^{2\Phi}} \ .
\end{equation}
It differs from the standard Natario--type metric~\eqref{C1} by the form factor
\begin{equation}
g(r) \ = \ \frac{e^\Phi}{\sqrt{1-\frac{b}{r}}} \ .
\end{equation}

In Cartesian coordinates this metric has the form
\begin{equation}\label{C3}
-dT^2+h_{ij}(dx^i+N^idT)(dx^j+N^jdT) \ ,
\end{equation}
with the intrinsic metric
\begin{eqnarray}
h_{ij} \ = \ \delta_{ij}+(g^2-1)\frac{x_ix_j}{r^2}\\ 
\left(h^{ij} \ = \ \delta_{ij}+(g^{-2}-1)\frac{x_ix_j}{r^2}\right)
\end{eqnarray}
and the shift vector
\begin{equation}
N^i=\frac{\beta}{g}\frac{x^i}{r} \ .
\end{equation}
After the embedding of the warp drive, the shift vector becomes
\begin{equation}
N^i=\left(1-f(\vec{r},T)\right)\frac{\beta}{g}\frac{x^i}{r}-f(\vec{r},T)v^i_s(T) \ .
\end{equation}
Since the lapse function $N$ is $1$, the metric~\eqref{C3} can also be considered a kind of warp drive metric, albeit different from the Natario--type ones (a warp drive with non--flat spatial slices has previously been proposed in~\cite{VanDenBroeck:1999sn}).

In order for the wormhole to be traversable, the metric has to satisfy three conditions: the flare--out condition
\begin{equation}
b(r) \ > \ rb'(r) \ \Leftrightarrow \ \left(\frac{1-\beta^2}{g^2}\right)' \ > \ 0 \ ,
\end{equation}
the throat condition
\begin{equation}
b(r_0) \ = \ r_0 \ ,
\end{equation}
and the absence of horizon condition
\begin{equation}
e^{2\phi} \ > \ 0 \ .
\end{equation}
From the two latter conditions it can be seen that the form factor $g$ has a singularity at $r=a$. To see if it's a real (curvature) or a coordinate singularity, let us compute the Ricci scalar. Without loss of generality, one can orient the $z$-axis along the velocity vector $\vec{v}_s$, so that $(\vec{v}_s\vec{r})=v_sr\cos\theta$. With this choice, the metric~\eqref{C3} can be written in spherical coordinates as
\begin{equation}\label{C4}
-(1-\Phi^2-\Psi^2)dT^2+2g\Phi dTdr+2\Psi rdtd\theta+g^2dr^2+r^2d\Omega^2 \ ,
\end{equation}
with
\begin{eqnarray}\label{K}
\Phi \ = \ (1-f)\beta-fgv_s\cos\theta \ , \ \Psi \ = \ fv_s\sin\theta \ .
\end{eqnarray}
In the simplest case when $f$ is not $\phi$-dependent (meaning that the warp drive is moving in the radial direction), the Ricci scalar is given by
\begin{equation}
R \ = \ \frac{1}{2r^2g^3}\left(Ag'+Bg'^2+Cg+Dgg'+Egg''+Fg^4+Gg^5+Hg^2+Ig^2g'+Jg^2g''+Kg^3+L\right) \ .
\end{equation}
Since the singularity in $g'$ scales like $g^3$, and the one in $g''$ scales like $g^5$, the coefficients in front of the divergent terms are the following:
\begin{eqnarray}
B \ = \ -E \ = \ 4r^2f(1-f)\beta v\cos\theta \ , \\
D \ = \ -4r^2v\cos\theta\partial_r\left((1-f)f\beta\right) \ , \\
F \ = \ 2\beta v\partial_\theta\left(f\cos\theta\right) \ , \\
G \ = \ (\partial_\theta\left(f\cos\theta\right))^2 \ ,\\
I \ = \  4fv^2r\cos\theta\left(f\cos\theta+3r\partial_rf\cos\theta-2\partial_\theta f\sin\theta\right) \ ,\\
J \ = \ 4r^2f^2v^2\cos^2\theta \ .
\end{eqnarray}
For a nonzero $f$, it's not possible to set all of them to zero, so the singularity is real. One may attempt to get rid of it by also interpolating the intrinsic metric:
\begin{equation}
h_{ij} \ = \ \delta_{ij}+(1-f)(g^2-1)\frac{x_ix_j}{r^2} \ ,
\end{equation}
resulting in the line element
\begin{equation}
-(1-\frac{G}{g^2}\Phi^2-\Psi^2)dt^2+2\frac{G}{g}\Phi dtdr+2\Psi rdtd\theta+Gdr^2+r^2d\Omega^2 \ .
\end{equation}
with $\Phi$ and $\Psi$ given by~\eqref{K}, and
\begin{equation}
G \ = \ (1-f)g^2+f \ .
\end{equation}
\begin{figure}
    \centering
    \includegraphics[width=0.7\linewidth]{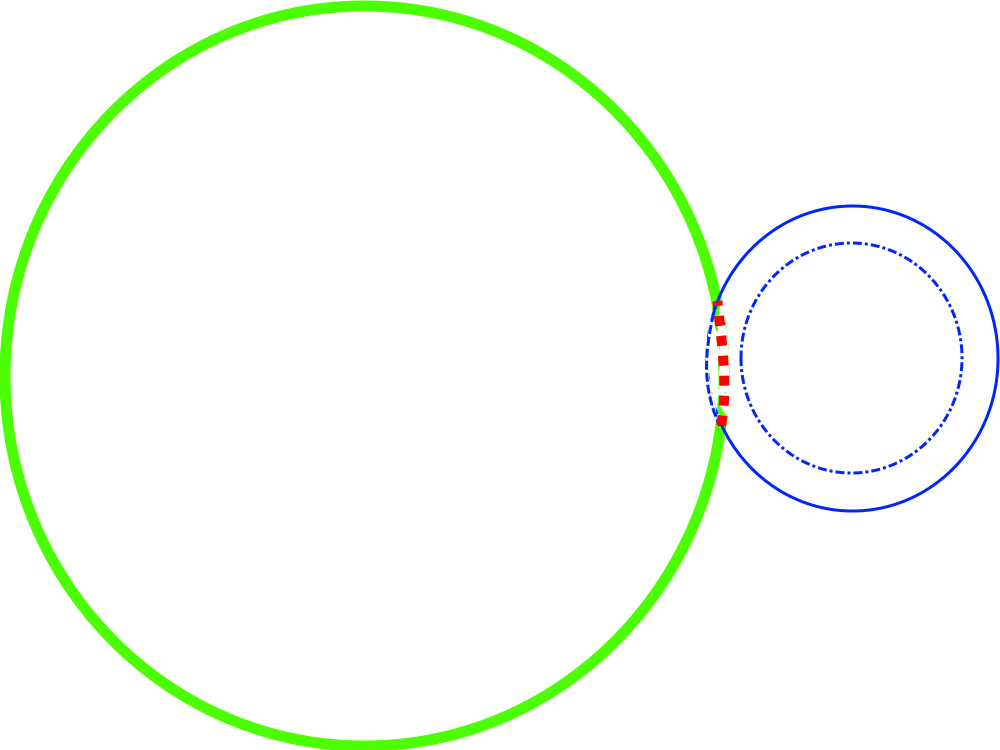}
    \caption{A warp drive bubble (thin solid blue line) traversing a wormhole (thick solid green line). Thin dashed blue line: the part of the warp drive on the other side of the wormhole. Thin dash--dotted blue line: the inner part of the bubble where $f=1$. In the region of the throat passing through the shell of the warp drive where $0<f<1$ (thick dotted red line), the singularity is not removable by coordinate transformation if the wormhole has no horizon.}
    \label{fig1}
\end{figure}
In that case, the metric inside the bubble would be locally Alcubierre; however, as the 00--component of the metric contains the terms
\begin{equation}
g^2(1-f)f^2v^2_s\cos^2\theta+2g(1-f)^2f\beta v_s\cos\theta \ ,
\end{equation}
the gravitational singularity would still be "felt" by the shell of the warp drive, i. e. the region where $0<f<1$ (fig.~\ref{fig1}).

Finally, one can try to remove the singularity by performing the coordinate transformation:
\begin{equation}\label{PL}
\rho(r) \ = \ \rho_0 \ + \ \int^r_{r_0} \ dr' \ \frac{g(r')}{\lambda(r')} \ ,
\end{equation}
where $\lambda$ is an arbitrary function that is finite and nonzero for $r_0\le r<\infty$, and tends to 1 in the limit $r\rightarrow\infty$. In particular, the choice
\begin{equation}
\lambda \ = \ e^{\Phi} \ , \ \rho_0 \ = \ 0
\end{equation}
yields the standard definition of proper length.

This would produce the line element
\begin{equation}
-dT^2+(\lambda(\rho)d\rho-\beta(\rho)dT)^2+r^2(\rho)d\Omega^2 \ ,
\end{equation}
but if we write it in Cartesian coordinates (defined as $x=\rho\sin\theta\cos\phi, y=\rho\sin\theta\sin\phi, z=\rho\cos\theta$) in the form~\eqref{C3}, it would have
\begin{eqnarray}\label{C2}
h_{ij} \ = \ \frac{r^2(\rho)}{\rho^2}\delta_{ij}+\left(\lambda^2(\rho)-\frac{r^2(\rho)}{\rho^2}\right)\frac{x_ix_j}{\rho^2} \ , \ N^i \ = \ \frac{\beta(\rho)}{\lambda(\rho)}\frac{x^i}{\rho} \ .
\end{eqnarray}
Then, after we embed the warp drive by changing $N^i$ to
\begin{equation}
N^i \ = \ (1-f)\frac{\beta}{\lambda}\frac{x^i}{\rho}-fv^i_s \ ,
\end{equation}
we obtain
\begin{equation}
g_{00} \ = \ -1+(1-f)^2\beta^2+2(1-f)f\lambda\beta fv\cos\theta+f^2v^2\left(\lambda^2\cos^2\theta+\frac{r^2}{\rho^2}\sin^2\theta\right) \ .
\end{equation}
In the limit $r\rightarrow r_0$ (i. e. $\rho\rightarrow\rho_0$), the last term would be dependent on the unphysical cutoff $\rho_0$, and if we set $\rho_0$ to zero, $g_{00}$ would once again be singular.

Hence we reach a paradoxical conclusion: \textit{if a wormhole has a horizon, it would not be "felt" inside the warp drive bubble, and the bubble can traverse it. And, vice versa, if a wormhole is humanly traversable, a warp drive would encounter a singularity while approaching it.}

One can also consider spherical warp drives, such as the one described in~\cite{Eroshenko:2022nph}: this class of solutions no longer has the physical interpretation of a localized "bubble" that can host a spacecraft, and may instead be understood as an incoming or outgoing spherical wave, similar to water ripples from a rock. Namely, if one starts with a metric of the form~\eqref{C2} and embeds a spherically symmetric warp drive:
\begin{equation}
N^i \ = \ (1-f)\frac{\beta}{\lambda}\frac{x^i}{\rho}-fv_s\frac{x^i}{\rho}
\end{equation}
with the shape function $f\left(r-r_s(t)\right)$ dependent only on the radial coordinate, we get
\begin{equation}
g_{00} \ = \ -1+\left((1-f)\beta-f\lambda v_s\right)^2 \ .
\end{equation}
Hence the resulting metric is singularity--free, but still dependent on the arbitrary function $\lambda$.

\section{Conclusions}\label{sec:conclusions}
In this paper, we introduced a particular coordinate transformation for the Morris--Thorne wormhole that is analogous to Painlevé--Gullstrand coordinates for the Schwarzschild black hole. Just like the Painlevé--Gullstrand coordinates were shown in~\cite{Ellis:2004aw} to correspond to a Natario--type warp drive, our transformation brings the wormhole metric to a different kind of warp drive metric that has nonzero intrinsic curvature. We analysed the traversability conditions and proved that whenever the wormhole is traversable, the warp drive would encounter a physical singularity at the throat, so, paradoxically, warp drives can traverse wormholes with horizons, but not humanly traversable ones. 

One should note, however, that there are three possible loopholes in this "no-go" theorem. First, one can avoid the singularity by applying the coordinate transformation~\eqref{PL}: while the parameters $\rho_0$ and $\lambda(r)$ are unphysical at first glance, they may be attributed physical meaning through the right--hand side of the Einstein equations, as solutions corresponding to a particular configuration of the stress--energy tensor.

Second, one can apply the procedure that is referred to as "surgery" in~\cite{Visser:1989kg}: namely, introduce a cutoff radius $a>r_0$, and then "glue together" the two ends of the wormhole at $r=a$ (i. e. relax the traversability condition that $b(r)=r$ at the minimal radius). This procedure would produce a wormhole with a sharp transition between the two ends, similar to the one described in~\cite{James:2015ima}.

Finally, the presence of a singularity does not necessarily mean that a spacetime is geodesically incomplete, and more thorough analysis is needed to see whether the geodesics are indeed discontinuous at the intersection between the warp drive's shell and the wormhole's throat.

All in all, further work is required to firmly establish whether humanly traversable wormholes are indeed impenetrable to warp drives, or there is some different lesson to learn from the singularities that emerge.

\acknowledgments %
We are grateful to Harold "Sonny" White for the discussions and for his useful feedback on earlier versions of this work, and to Prof. Claudio Maccone for his instructive comments and questions. The work is supported by the 2023 LSI grant "Traversable Wormholes: A Road to Interstellar Exploration". Part of the computations in this work was done with OGRe, a General Relativity Mathematica package developed by Barak Shoshany~\cite{Shoshany:2021iuc}.


\bibliographystyle{JCAP}

\end{document}